# A Generic Algorithm for Universal TDM Communication Over Inter Satellite Links

Miroslav Popovic
*University of Novi Sad*
*Faculty of Technical Sciences*
Novi Sad, Serbia
miroslav.popovic@rt-rk.uns.ac.rs

Marko Popovic
*RT-RK Institute for Computer Based Systems*
Novi Sad, Serbia
marko.popovic@rt-rk.com

Pavle Vasiljevic
*University of Novi Sad*
*Faculty of Technical Sciences*
Novi Sad, Serbia
pavle.vasiljevic@uns.ac.rs

Ilija Basicevic
*University of Novi Sad*
*Faculty of Technical Sciences*
Novi Sad, Serbia
ilija.basicevic@rt-rk.uns.ac.rs

***Abstract*—The original Python Testbed for Federated Learning Algorithms is a light FL framework, which provides the three generic algorithms: the centralized federated learning, the decentralized federated learning, and the TDM communication (i.e., peer data exchange) in the current time slot. The limitation of the latter is that it allows communication only between pairs of network nodes. This paper presents the new generic algorithm for the universal TDM communication that overcomes this limitation, such that a node can communicate with an arbitrary number of peers (assuming the peers also want to communicate with it). The paper covers: (i) the algorithm's theoretical foundation, (ii) the system design, and (iii) the system validation. The main advantage of the new algorithm is that it supports real-world TDM communications over inter satellite links.**

## I. Introduction

This research was conducted within the ongoing EU Horizon 2020 project TaRDIS [1] aiming to create a toolbox for easy programming of a broad range of distributed swarm applications, from smart grids, homes, and cities to robotics in Industry 4.0, and to navigation of low Earth orbit (LEO) satellite constellations. The edge system considered in this paper is a satellite constellation where satellites use federated learning to aid their navigation.

The Python Testbed for Federated Learning Algorithms (PTB-FLA) [2] is a light framework for federated learning algorithms (FLAs), which is written in pure Python to be easy to install and to fit to a small IoT's memory footprint. To aid easy programming in tune with low-code/no-code initiative, PTB-FLA offers: (i) a simple API that is amenable both to nonprofessional developers and LLMs such as ChatGPT, (ii) the 4-phase development paradigm for humans [3], and (iii) the adapted 4-phase and 2-phase development paradigms for ChatGPT [4].

The MicroPyton [5] Testbed for Federated Learning Algorithms (MPT-FLA) is a PTB-FLA derivative that inherits the PTB-FLA advantages and enables running MPT-FLA applications in fully distributed settings, such that individual application instances may run on different network nodes like PCs and IoTs in edge systems. Both frameworks are publicly available at [6], and since evolutionary changes of both are well aligned, we will keep focus on PTB-FLA in this paper.

The original PTB-FLA API is based on the Single Program Multiple Data (SPMD) pattern, and it provides the three generic algorithms: (i) the centralized federated learning, (ii) the decentralized federated learning, and (iii) the Time Division Multiplexing (TDM) communication (i.e., peer data exchange), in the current time slot, that may be used e.g., for orbit determination and time synchronization (ODTS) in LEO satellite constellations [7], [8], [9].

The limitation of the generic algorithm for TDM communication is that it allows communication only between pairs of network nodes. This support for pairwise communication is sufficient in the case of single antenna satellites but is insufficient in the case satellites have more antennas as they are then able to communicate with more peers simultaneously. This paper presents the new generic algorithm for the universal TDM communication that overcomes this limitation, such that a node can communicate with an arbitrary number of peers (assuming the peers also want to communicate with it).

Although the new generic algorithm is a rather simple and straightforward generalization of a previous one, it turned out to be surprisingly universal as it supports TDM communication among satellites where an individual satellite has (i) an arbitrary number of antennas (note that different satellites may have different numbers of antennas) and (ii) an arbitrary number of peers (note that the number of peers is less or equal to the number of antennas). To describe this universality more clearly, in this paper, we model the PTB-FLA application as a set of its instances and the universal TDM communication (i.e., data exchange) as an algebraic relation on this set.

After analyzing the algebraic properties of the new generic algorithm, we present the design of the new PTB-FLA version (which comprises the new algorithm) with focus on the algorithm's pseudocode and its operation. Finally, we present the new PTB-FLA version validation.

In summary, the main original paper contributions are: (1) the new generic algorithm's theoretical foundation, (2) the new PTB-FLA version design, and (3) the new PTB-FLA version validation. The paper is organized as follows. Section II presents the theoretical foundation, Sections III and IV present the new PTB-FLA version design and validation, respectively, and Section IV concludes the paper.

## II. Theoretical Foundation

In this section, we model the PTB-FLA application as a set of its instances and the universal TDM *communication* (i.e., *data exchange*; note that we use these terms interchangeably) as an algebraic relation on this set. We start with the definition of this relation and then we analyze its properties. To save space, we assume that readers are familiar with relevant definitions and theorems from discrete mathematics e.g., [10].

Let A = $\{a_1, a_2, \ldots, a_m\}$, $m \leq n$, be a set of application instances participating in the TDM data exchange in the current time slot. The collective TDM data exchange among application instances is a relation $R$ on $A$ i.e., $R$ is a subset of Cartesian product $A$ x $A$. If $(a_i, a_j)$ is in $R$, we write $a_iRa_j$ and say that $a_i$ is related to $a_j$ by means of $R$, or simply $a_i$ is related to $a_j$.

The semantic of $R$ is exchanging data i.e., $aRb$ means that $a$ sends its data to $b$ and receives $b$'s data from $b$ (imagine $a$ and $b$ having two hands – with the left they give their data and with the right they get the peer's data). Obviously, data exchange between $a$ and $b$ is only possible if both $aRb$ and $bRa$ are in $R$. An example of the simplest possible $R$ is $R_1 = \{(a, b), (b, a)\}$.

In a more complex relation $R$, an application instance can exchange data with more than one instance i.e., it may have more peers – in the "hand pairs" metaphor this means that it has as many pairs as peers i.e., a pair of hands per peer e.g., in $R_2 = \{(a, b), (b, a), (b, c), (c, b)\}$, an instance $b$ simultaneously exchanges data with the instance $a$ and the instance $c$, while $a$ and $c$ only exchange data with $b$. Here $b$ has two pairs of hands whereas $a$ and $c$ have a single pair each.

In satellite communications is perhaps rarely the case, but each instance can exchange data with all other instances e.g., $R_3 = \{(a, b), (b, a), (a, c), (c, a), (b, c), (c, b)\}$. Here each instance has a pair of hands for each (other) instance.

Relation $R$ has the following five properties, which are easy to prove as they directly follow from the properties' definitions (see e.g., [10]).

### A. Property 1 (Inverse Relation)

$R^{-1} = R$, that is the inverse relation of $R$, denoted by $R^{-1}$, is equal to $R$.

### B. Property 2 (Data Propagation)

The composition of $R$ relations leads to data propagation. Consider the following example.

Let $R_{21} = \{(a, b), (b, a)\}$, $R_{22} = \{(b, c), (c, b)\}$.

Then $R_{21} \circ R_{22} = \{(a, c)\}$ and $R_{22} \circ R_{21} = \{(c, a)\}$.

Note that these compositions are not $R$ relations, but their union is a valid $R$ relation $R_{23} = R_{21} \circ R_{22}$ U $R_{22} \circ R_{21} = \{(a, c), (c, a)\}$. Here data propagates from $a$ over $b$ to $c$ and from $c$ over $b$ to $a$.

Furthermore, (based on the well-known theorem) composition of relations is associative i.e., when evaluating a sequence of $R$ compositions from left to right any grouping of individual relations is allowed. As the result of $R$ compositions, data of application instances participating in the leftmost relation may propagate to the instances participating in the rightmost relation. Symmetrically, in a reverse sequence of $R$ compositions, data may propagate from the instances participating in the rightmost composition (of the original sequence) to the application instances participating in the leftmost composition.

### C. Property 3 (Special Properties)

According to the definitions of relation reflexivity, symmetricity, transitivity, and anti-symmetricity, $R$: (1) is not reflexive, (2) is symmetric, (3) is not transitive, and (4) is not anti-symmetric.

### D. Property 4 (Symmetric Closure)

By the definition of the symmetric closure of $R$, $R$ is its own symmetric closure.

### E. Property 5 (Graph Representation)

Since $R$ is a symmetric anti-reflexive relation, it may be represented by a graph $G(V, E)$, where $V = A$, and $E$ is a collection of two-element subsets of $A$ defined by $\{a, b\}$ is in $E$ if and only if $(a, b)$ is in $R$ (note that $(b, a)$ also must be in $R$).

Here are three examples of $R$ relations (see Fig. 1):

For $R_1$, $E_1 = \{\{a, b\}\}$, see Fig. 1(a).

For $R_2$, $E_2 = \{\{a, b\}, \{b, c\}\}$, see Fig. 1(b).

For $R_3$, $E_3 = \{\{a, b\}, \{a, c\}, \{b, c\}\}$, see Fig. 1(c).

(a)

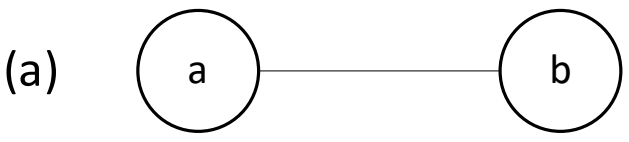

(b)

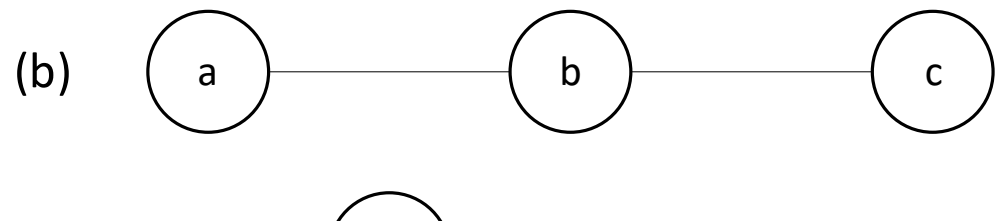

(c)

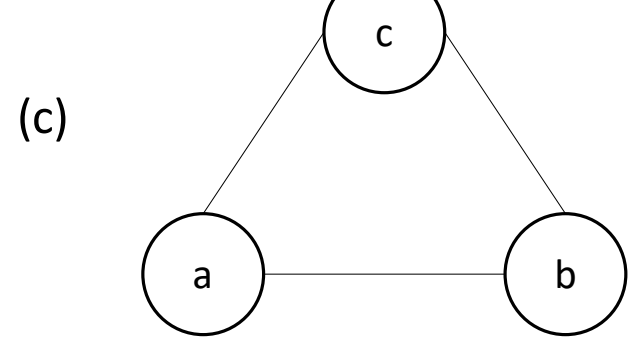

Fig. 1. The examples of $R$ relations.

This concludes with the presentation of $R$'s properties. Note that $R$ is just a relation and is not some of the higher algebraic structures like partial ordering, total ordering, etc.

## III. PTB-FLA Design

This section presents the PTB-FLA design details. The next subsections present the system architecture and the system operation, respectively.

### A. PTB-FLA System Architecture

The PTB-FLA-based system, briefly called the PTB-FLA system, is a distributed system that may be represented as a graph with $n$ nodes (with the node IDs from 1 to $n$, respectively), which are interconnected with edges, where the nodes are processes hosted by processors (e.g., IoTs, smart devices, computers, etc.) and the edges are communication links (TCP connections), see Fig. 2. The edges in Fig. 2 are shown as dashed lines to reflect the fact that the communication links are dynamically created and destroyed.

The set of links that are created depends on the system operation mode i.e., on the generic algorithm executed by PTB-FLA. When visualizing the current operation mode, we only draw the links that are created for that mode, so the abstract graph shown in Fig. 2 materializes into: (1) a star in the case of the generic centralized FLA, (2) a clique i.e., a complete graph in the case of the generic decentralized FLA, and (3) a *R* relation (see Section II) in the case of the universal TDM communication i.e., peer data exchange in the current time slot.

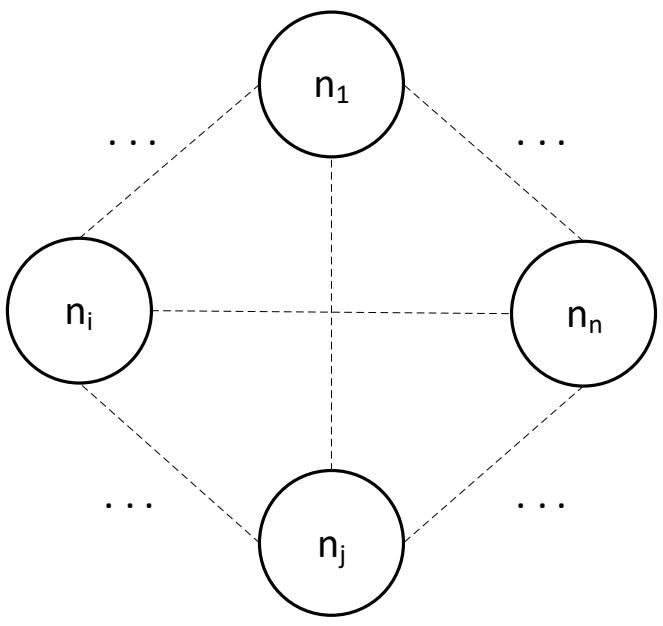

Fig. 2. The block diagram of the PTB-FLA system architecture.

Generally, a node $n_i$ is a process executing an application instances $a_i$. Each $a_i$ initially creates its testbed instance $t_i$. A set of all the application instances constitutes a distributed application $A = \{a_1, \ldots, a_n\}$, whereas a set of all the testbed instances constitutes a distributed testbed $T = \{t_1, \ldots, t_n\}$. During normal system operation, the distributed application *A* uses the distributed testbed *T* to execute the desired distributed algorithm by using the appropriate generic distributed algorithm offered by PTB-FLA. New PTB-FLA version offers: (1) the generic centralized FLA, (2) the generic decentralized FLA, and (3) the new generic universal TDM communication algorithm. More information on the original PTB-FLA architecture is available in prior works [1], [6].

In the new PTB-FLA version, the original PTB-FLA API [2] was extended with the function getMeas that implements the new generic algorithm. The signature of the function getMeas is the following:

*obss* getMeas(*peer_ids, odata*)

Where *peer_ids* is the list of an arbitrary number $k$ ($k > 0$) of peer node IDs for all the nodes this node communicates with, the *odata* is the orbital data of this node, and the return value *obss* is the list of orbital data received from the peer nodes where each element of the list *obss* corresponds to the element in the same position of the list *peer_ids*.

The two important assumptions of the function getMeas are the following: (a) if a node takes part in the communication in the given time slot, it can talk an arbitrary number of peer nodes in that time slot, of course, under condition that all the peer nodes also want to communicate with this node, (b) if a node does not take part in the communication in some time slot, then it should skip that time slot by calling getMeas and setting *odata* to None.

### B. PTB-FLA System Operation

The main four PTB-FLA operation modes are: (i) the system startup, (ii) the generic centralized FLA execution, (iii) the generic decentralized FLA execution, and (iv) the TDM communication supported by the function getMeas. Here we present the last one (for more info on the first three see [2]).

The function getMeas, see Algorithms 1, has two arguments: *peerIds* and *odata* (line 2), which are the list of the peer identifications and the data this node wants to send to its peers, respectively, and it returns the list *peerOdatas* that contains data received from the peers (line 28). Besides the arguments, the function getMeas operates on the following PTB-FLA instance data: *nodeId*, *timeSlot*, *timeSlotsMap*, which are the node identification, the current time slot, and the buffer of messages sent by faster peer nodes sent to this node (from the next time slots), respectively.

In the beginning, getMeas checks whether this node wants to skip the current time slot, by checking whether *odata* is set to None (see line 4), and if yes, then it just increments the *timeSlot* (line 5) and returns None (see line 6).

Algorithm 1. The function getMeas.

```
01: // PTB-FLA instance data: nodeId, timeSlot, timeSlotsMap
02: def getMeas(peerIds, odata):
03:   // If odata is None, this node wants to skip this time slot
04:   if odata == None:
05:     timeSlot += 1  // Increment time slot
06:     return None
07:   // Send own odata to the peers and then receive peers odata
08:   for peerId in peerIds
09:     sendMsg(peerId, [timeSlot, nodeId, odata])
10:   peerOdatas = []
11:   for peerId in peerIds:
12:     if (timeSlot, peerId) in timeSlotsMap
13:       msg = timeSlotsMap[(timeSlot, peerId)]
14:       del timeSlotsMap[(timeSlot, peerId)]
15:     else
16:       while True
17:         msg = rcvMsg()
18:         peerTimeSlot, peerNodeId, peerOdata = msg
19:         if (peerTimeSlot, peerNodeId) != (timeSlot, peerId)
20:           timeSlotsMap[(peerTimeSlot, peerNodeId)] = msg
21:           continue
22:         else
23:           break
24:     // Unpack msg and add peerOdata to peerOdatas
25:     peerTimeSlot, peerNodeId, peerOdata = msg
26:     peerOdatas.append(peerOdata)
27:   timeSlot += 1  // Increment time slot
28:   return peerOdatas
```

Otherwise, i.e., if this node takes part in the current time slot), then it first sends its data to all its peers (lines 8-9), and then it receives data from all its peers (lines 10-26). Receiving data is a bit involved, because some of its peers may be faster and may have already sent their messages for the next time slots. Therefore, this node first checks whether there are such messages stored in *timeSlotsMap*, and if yes it retrieves them from there (lines 12-14).

If there are no such messages, then getMeas receives a new message (line 17), and unpacks it into the individual message fields (line 18). Next, getMeas checks whether the new message is related to the current time slot (line 19). If not i.e., if the message is from the next time slot, getMeas stores it into *timeSlotsMap* (line 20), and it continues receiving subsequent messages (line 21). If the message is related to the current time slot, then getMeas unpacks the message (line 25), and adds the received peer's data to the list *peerOdatas* (line 26).

Finally, getMeas increments the current time slot (line 17), and returns peers data collected in the list *peerOdatas* (line 28).

## IV. PTB-FLA System Evaluation

In order to assess the performance of the PTB-FLA system and its generic TDM communication capabilities, we developed a dedicated benchmarking application designed to simulate communication within a clique of nodes, which we consider to be the worst case scenario for this type of communication. In a real physical system this would be equivalent to a satellite constellation where every satellite has a direct link to all others. This could be done either via a pairwise links, facilitating pairwise communication between satellites, or by using multiple communication links simultaneously (i.e. by having multiple antennas pointed at each other). Within PTB-FLA we developed TDM communication primitives to cover both cases, get1meas for the pairwise links, and getMeas, as presented in this paper, for using multiple links simultaneously. The benchmarking application runs on a single host (i7-8550u 16GB of RAM), which repeatedly runs the two testing applications. For get1meas we generated the schedule as a round robin tournament, resulting in a deterministic map of communication inside time slots for every node, while for getMeas the schedule was given as a list of all other node IDs. These two test applications are semantically equivalent and output the same result, exchanging data between all nodes albeit achieving it in different ways. During system evaluation, the two test applications were run with node counts ranging from 20 to 200 in increments of 20, with each configuration executed 50 times. Execution times for the semantically equivalent sections of code were measured and stored within the evaluation database for each node of each run. Evaluation results were then subsequently grouped by the configuration that produced them and averaged out.

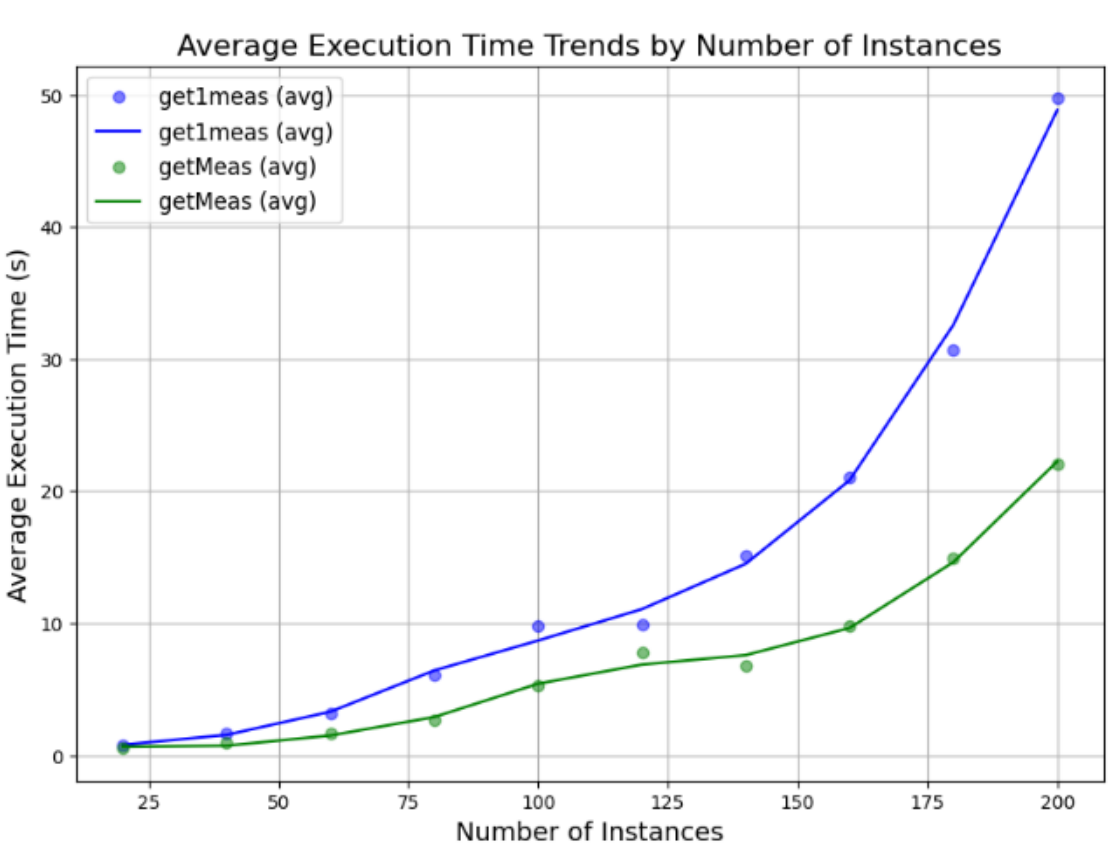

Fig. 3. Average execution time versus the number of nodes

For both scenarios, the average execution time was expected to grow quadratically, which would be consistent with the O($n^2$) growth of the number of edges in a clique as the number of vertices increase. As expected both functions exhibit this behavior, with get1meas showing growth that is faster by a constant from getMeas. Results are shown on Fig 3. with the upper line representing the average execution time of get1meas and the lower line representing the average execution time of getMeas respectively.

## V. Conclusion

This paper presents the new generic algorithm for the universal TDM communication, such that a node can communicate with an arbitrary number of peers (assuming the peers also want to communicate with it) in the current time slot.

The main advantage of the new algorithm is that it supports real-world TDM communications over inter satellite links, as it supports TDM communication among satellites where an individual satellite has (i) an arbitrary number of antennas (note that different satellites may have different numbers of antennas) and (ii) an arbitrary number of peers.

To the best of authors' knowledge, the new algorithm has no limitations. In the future we plan to conduct PTB-FLA experimental evaluation using different topologies and as a part of more complex and dynamic systems.

## Acknowledgment

Funded by the European Union (TaRDIS, 101093006). Views and opinions expressed are however those of the author(s) only and do not necessarily reflect those of the European Union. Neither the European Union nor the granting authority can be held responsible for them.